\documentclass[10pt,secnumroman]{article}

\usepackage{amsmath}
\usepackage{amssymb}
\usepackage{graphics}
\usepackage{rotating}
\usepackage{cite}
\usepackage{color}
\usepackage{fancybox}
\usepackage{pstricks}
\usepackage{bbm}
\usepackage{bm}
\usepackage{epsfig}

\def\0{\mbox{\tiny $0$}}
\def\1{\mbox{\tiny $1$}}
\def\2{\mbox{\tiny $2$}}
\def\3{\mbox{\tiny $3$}}
\def\4{\mbox{\tiny $4$}}
\def\5{\mbox{\tiny $5$}}
\def\6{\mbox{\tiny $6$}}
\def\7{\mbox{\tiny $7$}}
\def\8{\mbox{\tiny $8$}}
\def\9{\mbox{\tiny $9$}}
\title{\shadowbox{\large \bf DIRAC SOLUTIONS FOR QUATERNIONIC POTENTIALS }}

\author{
\small  Stefano De Leo\thanks{deleo@ime.unicamp.br} \,\,
and\, Sergio Giardino\thanks{giardino@ime.unicamp.br}\\
\small Department of Applied Mathematics,
State University of Campinas, Brazil}

\date{\small
\fcolorbox{black}{yellow} {\color{red} $\bullet$ {\color{black}{
{\footnotesize  {\sc Journal of Mathematical Physcis} {\bf 55}, 022301-10 (2014) }}}
{\color{red}{$\bullet$}} } }

\begin{document}
%

\maketitle

\vspace*{-.7cm}

\begin{abstract}
\noindent
The Dirac equation is solved for quaternionic
potentials, $i\,V_{\0}+j\,W_{\0}$ ($V_{\0}\in \mathbb{R}\,,\,\,W_{\0}\in \mathbb{C}$). The study shows two different solutions. The first one contains  particle
and anti-particle solutions and leads to  the diffusion, tunneling and Klein energy
zones. The standard solution  is recovered taking the complex limit of this solution.  The second solution, which does not have a complex counterpart, can be seen as a $V_{\0}$\,-\,antiparticle or $|W_{\0}|$\,-\,particle solution.
\end{abstract}

\maketitle
\section{ \normalsize  INTRODUCTION}\label{s1}

Analytical solutions are very important in physics. They
permit, by a detailed analysis of their mathematical structure, a conceptual understanding
of the physical problem and, more important, allow to analyze  more complex solution by
approximations of analytical cases. Complex models are often studied by using
simple systems as building blocks and  reduce to simple models in the appropriate limit\cite{Cohen}.
 One of the simplest and important model in relativistic quantum mechanics is the
one-dimensional step\cite{Zuber,Sakurai,Gross}
\begin{equation*}
V(z)=\left\{\;
V_0\mbox{ for } z>0\; \mbox{ and }\; 0\mbox{ for } z<0\;\right\}.
\end{equation*}
Depending on the energy $E$ of the incoming  particle, the wave-function has either
oscillating or evanescent solutions,
\begin{equation*}
\begin{array}{llrcccl}
\mbox{\bf D)} & \mbox{oscillating particles diffusion zone:} &  & & E & > &  V_{\0}+m\,\,,\\
\mbox{\bf E)} & \mbox{evanescent waves zone:} & \mbox{Max}\left[\,m\,,\, V_{\0}-m \,\right]&<&E&<&V_{\0}+m\,\,,\\
\mbox{\bf K)} & \mbox{oscillating antiparticles Klein zone:} & m&<&E&<&\mbox{Max}\left[\,m\,,\, V_{\0}-m\,\right]\,\,.
\end{array}
\end{equation*}
The simplest cases of quantum scattering come from
studying this potential. In the Klein zone, the reflection coefficient becomes greater than one and this is interpreted as a pair production in the potential region. The oscillating solutions in this region are thus identified as antiparticles\cite{Klein1,Klein2,Klein3,Klein4}. For thin barrier ($V_{\0}$  restricted to
a small finite region of space) the evanescent waves propagates from the left to the right boundary of the barrier and produce the well known phenomenon of quantum tunnelling.
Particles diffusion and quantum tunneling  are valid  for non-relativistic quantum mechanics
described by the Schr\"{o}din\-ger equation\cite{Cohen} and for relativistic
quantum mechanics described by the Dirac equation\cite{Zuber,Sakurai,Gross}. The Klein pair production is a clear relativistic effect. The scattering of the Dirac field by the electrostatic potentials, like step and  barrier, has been the subject matter of  recent investigations in complex quantum mechanics\cite{Com1,Com2,Com3,Com4,Com5,Com6,Com7,Com8,Com9}. In the context of quaternionic quantum mechanics\cite{Adler},
pioneering\cite{ref1989} and recent works on the  quaternionic non-relativistic
scattering\cite{NR1,NR1r,NR2,NR3,NR4,NR5,NR5r,NR6,NR7,NR7r} and on the quaternionic Dirac
fields\cite{Rel0,Rel1,Rel1r,Rel2,Rel3,Rel4,Rel5,Rel6,Rel7,Rel8,Rel8r,Rel9,Rel9r,Rel10} have been outlined the possibility to see qualitative and quantitative difference between the complex and quaternionic formulation of quantum mechanics.

In this article, the Dirac equation is solved for quaternion
potentials in a one-dimension space. In a quaternionic scattering, a complex wave-function interacts with a quaternionic
potential, and quaternionic reflected and transmitted wave-function are generated. In the limit where the pure quaternion part of the potential goes to zero, the complex
Dirac wave-function has to be recovered.
The solutions obtained in this paper are basic for tackling more
complex phenomena of relativistic quaternionic scattering and tunneling in future studies.

This article is organized as follows: in Section \ref{s2}, we briefly introduce the
quaternionic Dirac equation;  in Section
\ref{s3}, the matrix equations are solved and the momenta, in the quaternionic potential region, are calculated (their behavior  allows to determined the {\em new} diffusion, tunneling
and Klein energy zones); Section \ref{s4} presents a discussion about
particles and anti-particles by a detailed analysis of the velocities; in Section \ref{s5}, we give the explicit quaternionic wave-functions and  analyze the limit cases; our conclusions and a discussion of future investigations are drawn in Section \ref{s6}.

\section{\normalsize THE QUATERNIONIC DIRAC EQUATION}\label{s2}

A complex anti-self-adjoint Hamilton operator can be expressed in terms of a self-adjoint operator $\hat{\mathcal{H}}$ by writing
$\hat{\mathcal{A}}=i\,\hat{\mathcal{H}}$.
For a quaternion Hamiltonian this in general cannot be done, because
\begin{equation}
\hat{\mathcal{A}}^\dagger=\big(i\,\hat{\mathcal{H}}\big)^\dagger=-\hat{\mathcal{H}}\,i\neq -\hat{\mathcal{A}}\,\,.
\end{equation}
Then, in quaternionic quantum mechanics anti-self-adjoint
operators are  usual in order to simplify the calculations. Following this
convention\cite{Adler}, an anti-hermitian Hamiltonian
operator is used in the Dirac equation for an arbitrary quaternionic
potential in the natural system of units, $\hbar=c=1$, so that
\begin{equation}\label{dir_eq}
\partial_t\,\Psi(\bm
r,\,t)=-\Big[\bm{\alpha\cdot\nabla}+i\,\beta\,m+\bm{h\cdot V}(\bm r) \Big]\,\Psi(\bm r,\,t)\,\,,
\end{equation}
where $V_s(\bm r)$ for $s=\{1,2,3\}$ are the real components of $\bm V$ and
the complex unities are contained in $\bm h=(i, j, k)$. The matrices
$\bm{\alpha}$ and $\beta$ satisfy the algebra
\begin{equation}
\bm{\alpha}=\bm{\alpha}^\dagger\,,\qquad\beta=\beta^\dagger\,,
\qquad\bm{\alpha}^2=\beta^2=1\,,
\qquad\{\,\beta\,,\,\bm{\alpha}\,\}=0\qquad\mbox{and}\qquad\{\alpha_r,\,\alpha_r\}=2\,\delta_{rs}\,\mathbbm{1};
\end{equation}
and the representation adopted for the calculations is the Dirac representation\cite{Zuber}
\begin{equation}
\bm{\alpha}=\left(
\begin{array}{cc}
0&\bm{\sigma}\\
\bm{\sigma}&0\\
\end{array}\right),\qquad
\beta=\left(
\begin{array}{cc}
\mathbbm{1}&0\\
0&-\mathbbm{1}\\
\end{array}
\right)\,\,.
\end{equation}
 The anti-hermiticity of
$\bm{\nabla}$ is seen through integration by parts of  $\langle \Psi|
\bm{\nabla}\Phi\rangle$. The
anti-hermiticity of the other elements of Eq.\,(\ref{dir_eq}) are trivial,
hence the anti-hermiticity of the operator is guaranteed and the
calculations for one-dimensional time independent potentials can be carried on.
Let us then consider the step potential
\begin{equation}\label{pot}
\bm V(\bm r,\,t)=\left\{\;
\bm V\mbox{ for } z>0\; \mbox{ and }\; 0\mbox{ for } z<0\;\right\}.
\end{equation}
with $\bm{V}$ a real constant vector. Using the function,
\begin{equation}\label{dea}
\Psi(\bm r,\,t)=\psi \,\exp[\,i\,(\,Q\,z\,-\,E\,t\,)\,]\,\,,
\end{equation}
where $\psi$ is a quaternionic constant spinor, the Dirac equation (\ref{dir_eq}) becomes
\begin{equation}\label{dirac_0}
-\psi\,i\,E+\alpha_{\3}\,\psi\,i\,Q+\big(\,i\,\beta\,m\,+\,\bm{h\cdot V}\,\big)\,\psi=0\,\,.
\end{equation}
By introducing $V_{\0}=V_{\1}$ and $W_{\0}=V_{\3}+i\,V_{\2}$,
\[\bm{h\cdot V}=i\,V_{\0}+k\,W_{\0}\,\,,\]
and multiplying Eq.\,(\ref{dirac_0}) by $i$
from the right, we obtain
\begin{equation}\label{dirac_00}
\big(\,E-\alpha_{\3}\,Q\,\big)\,\psi\,+\big(\,\beta\,m+V_{\0}-j\,W_{\0}\big)i\,\psi\,i=0\,\,.
\end{equation}
In the next section, we shall study in detail the solution of Eq.\,(\ref{dirac_00}). In particular, we shall focus our attention on the momentum in the potential region and its properties.

\section{\normalsize DIFFUSION, TUNNELING AND KLEIN ENERGY ZONES}\label{s3}
The constant spinor $\psi$ may be written in a symplectic manner as\cite{Rel1,Rel2,Rel3,Rel4}
\begin{equation}\label{des}
\psi=u+j\,w
\end{equation}
where $u$ and $w$ are complex constant spinors. Hence,
Eq.\,(\ref{dirac_00}) is split into two coupled complex equations
\begin{equation}
\label{emms}
M_{_-}u=-\,W^*_{\0}\,w\qquad\mbox{and}\qquad M_{_{+}}\,w=-\,W_{\0}\, u,
\end{equation}
where
\begin{equation}
M_{_{-}}=E-\alpha_{\3}\,Q-\beta\,m-V_{\0}\qquad\mbox{and}\qquad M_{_+}=E-\alpha_{\3}\,Q+\beta\,m+V_{\0}\,\,.
\end{equation}
By using Eq.\,(\ref{emms}), we can easily separate the equations for $u$ and
$w$,
\begin{equation}
\label{eigen_u}
 M_{_+} M_{_-}u=|W_0|^{^2}u\qquad\mbox{and}\qquad M_{_-} M_{_+}w=|W_0|^{^2}w
\,\,,
\end{equation}
where
\begin{equation}\label{evw0}
M_{_+} M_{_-}= \left[\,M_{_-} M_{_+}\,\right]^{t} = \left[
\begin{array}{cc}
E^{^2}+Q^{^2}-(V_{\0}+m)^{^2} &-2\,Q\,(E+m)\,\sigma_{\3}\\
-2\,Q\,(E-m)\,\sigma_{\3}&E^{^2}+Q^{^2}-(V_{\0}-m)^{^{2}}
\end{array}
\right]\,\,.
\end{equation}
Not trivial solutions are obtained by imposing
\begin{equation}
\mbox{det} \big[\,M_{_+} M_{_-}-|W_{\0}|^{^2}\,\big]= \mbox{det} \big[\,M_{_-} M_{_+}-|W_{\0}|^{^2}\,\big]=0\,\,.
\end{equation}
The previous constraint implies
\begin{equation}
Q_{_{\pm}}^{^2}=q_{_\pm}^{\2}+|W_{\0}|^{^{2}}\pm 2\,\delta\,\,,
\label{momentum}
\end{equation}
with
\[ q_{_\pm}^{\2}=\big(\,E\pm V_{\0}\,\big)^{^2}-m^{\2}\qquad\mbox{and}\qquad
\delta=\sqrt{E^{^2}V_{\0}^{^2}+p^{\2}|W_{\0}|^{^2}}-E\,V_{\0}\,\,.
\]
The sign of the squared momenta given in Eq.\,(\ref{momentum}) defines the
physical character of the wave-function, either oscillating or
evanescent in the $z>0$ region. Oscillating solutions may be particles or
anti-particles, and they will be discussed in the next section. The positive
definite $Q_{_+}^{^2}$ defines oscillating solutions only. On the
other hand, $Q_{_-}^{^2}$ may be either positive or negative,
\begin{equation}
\begin{array}{lcl}
\mbox{\bf D)} & Q_{_-}^{^2}>0 & \mbox{for}\,\,\,E>\sqrt{|W_{\0}|^{^{2}}+(\,V_{\0}+m\,)^{^{2}}}\,\,,\\
\mbox{\bf E)} & Q_{_-}^{^2}<0 & \mbox{for}\,\,\,\mbox{Max}\left[\,m\,,\, \sqrt{|W_{\0}|^{^{2}}+(\,V_{\0}-m\,)^{^{2}}} \,\right]<E<\sqrt{|W_{\0}|^{^{2}}+(\,V_{\0}+m\,)^{^{2}}}\,\,,\\
\mbox{\bf K)} & Q_{_-}^{^2}>0 & \mbox{for}\,\,\,m<E<\mbox{Max}\left[\,m\,,\, \sqrt{|W_{\0}|^{^{2}}+(\,V_{\0}-m\,)^{^{2}}} \,\right]\,\,.
\end{array}
\end{equation}
The D (diffusion) zone is characterized by oscillating solutions. In the complex limit, these solution in the potential region  represent  particles\cite{Zuber,Sakurai,Gross}. The K (Klein) zone is also characterized by  oscillating functions. In the complex limit, such solutions represent antiparticles moving in the potential regions\cite{Klein1,Klein2,Klein3,Klein4}.  The E (evanescent) zone  has solutions going to zero with $z$. If the evanescent zone is
finite, particles can tunnel across it. The energy range of the {\em quaternionic tunneling} is
\begin{equation}\label{energyrange}
\Delta E=\sqrt{|W_{\0}|^{^{2}}+(\,V_{\0}+m\,)^{^{2}}} - \mbox{Max}\left[\,m\,,\, \sqrt{|W_{\0}|^{^{2}}+(\,V_{\0}-m\,)^{^{2}}}\,\right]\,\,.
\end{equation}
For $|W_{\0}|\,\,\to\,\,0$, we recover the energy range of the complex tunneling, i.e.
 \[
 \Delta E_{_{\rm complex}} =\left\{\begin{array}{cccr}
 V_{\0} & \hspace*{.5cm}  & \mbox{for}& 0<V_{\0}<2\,m\,\,,\\
  2\, m & & & V_{\0}>2\,m\,\,.
 \end{array}\right.
 \]
 For $|W_{\0}|^{^{2}}+(\,V_{\0}-m\,)^{^{2}} =m^{\2}$,
 dashed line in Fig.\,1, we find the following energy range
 \begin{equation}
\label{circle}
 \frac{\Delta E_{_{\rm circle}}}{m} = \sqrt{1+\,4\,\frac{V_{\0}}{m}} \,-\, 1\,\,.
\end{equation}
 The points on the dashed line can be thus characterized by their $x$ and $y$ position and the correspondent  tunneling energy range. For the curves drawn in Fig.\,1, we have
\begin{equation}
\left(\,\frac{V_{\0}}{m}\,,\,\sqrt{2\,\frac{V_{\0}}{m}\,-\,\frac{V^{^{2}}_{\0}}{m^{^{2}}}  }\,\,;\,\sqrt{1+\,4\,\frac{V_{\0}}{m}} \,-\, 1\,\right)\,=\,
\left\{\begin{array}{lcccccr}
( & 0.05 & , & 0.32 & ; & 0.10 &)\,\,,\\
( & 0.31 & , & 0.73 & ; & 0.50 &)\,\,,\\
( & 0.75 & , & 0.97 & ; & 1.00 &)\,\,,\\
( & 1.31 & , & 0.95 & ; & 1.50 &)\,\,,\\
( & 1.71 & , & 0.70 & ; & 1.80 &)\,\,,\\
( & 1.93 & , & 0.38 & ; & 1.95 &)\,\,.
\end{array}
\right.
\end{equation}
Inside the circle, see Fig.\,1,
\begin{equation}
\label{inside}
\Delta E_{_{\rm inside}}=\sqrt{|W_{\0}|^{^{2}}+(\,V_{\0}+m\,)^{^{2}}} \,-\, m\,\,.
\end{equation}
Consequently, the complex and pure quaternionic part of the potential have the same {\em qualitative} effect on the tunneling energy range. For a fixed pure quaternionic  (complex) potential,  by increasing the  complex (pure quaternionic) perturbation we get an increased  energy range. Due to the additional term $2\,m\,V_{\0}$, the variation is greater for complex than for pure quaternionic  perturbations, see Fig.\,1. Outside the circle, we have
\begin{equation}
\label{outside}
\Delta E_{_{\rm outside}}=\sqrt{|W_{\0}|^{^{2}}+\,(\,V_{\0}+m\,)^{^{2}}} \,-\,  \sqrt{|W_{\0}|^{^{2}}+\,(\,V_{\0}-m\,)^{^{2}}}\,\,.
\end{equation}
Considering small quaternionic perturbations on the complex potential $V_{\0}(>2\,m)$, we find
\[
\Delta E_{_{\rm outside}}\,\approx\, 2\,m \,-\,\frac{m\,|W_{\0}|^{^{2}}}{V_{\0}^{^{2}}-m^{\2}}\,\,.
\]
This implies a decreasing  tunneling range zone for increasing values of quaternionic perturbations. For small complex perturbations on the pure quaternionic potential $|W_{\0}|$, the energy range can be approximated by
\begin{equation}
\label{outside}
\Delta E_{_{\rm outside}}=\sqrt{|W_{\0}|^{^{2}}+\, m^{\2} + 2\,m\, V_{\0} } \,-\,  \sqrt{|W_{\0}|^{^{2}}+\, m^{\2} - 2\,m\, V_{\0}}\,\,.
\end{equation}
The energy range behavior is now completely different. Complex perturbations increase the tunneling energy zone, see Fig.\,1.

\section{\normalsize PARTICLES AND ANTIPARTICLES}\label{s4}

The velocity analysis permits us to identify the
nature of the propagation in the potential region. In particular, it is possible to distinguish between  particles and anti-particles propagation. In complex relativistic quantum mechanics, the velocity is given by
\[ v_{_-}\left[\,V_{\0}\,,\,0\,\right]=\Big(\frac{\partial q_{_{-}}}{\partial E}\Big)^{^{-1}}=\frac{q_{_-}}{E - V_{\0}} = \frac{\sqrt{(E-V_{\0})^{^{2}}-m^{\2}}}{E-V_{\0}}\,\,.\]
For over potential solutions, $E>V_{\0}$, we have a  positive velocity which decreases  by increasing the electrostatic potential. Thus,  such solutions represent particles moving, in the potential region,  from the right to the left. For below potential solutions, $E<V_{\0}$, we find a negative velocity and
by invoking the Feynman-Stuckelburg\cite{Zuber,Sakurai,Gross} rule, $V_{\0}\to-\,V_{\0}$, we identify
such solutions as antiparticles moving, in the potential region, from the right to the left. This interpretation require pair production\cite{Klein1,Klein2,Klein3,Com3}. In this case,
electrostatic potential accelerates the anti-particles.

 In quaternionic relativistic quantum mechanics,
we have the momentum $Q_{_{-}}$ which has in $q_{_{-}}$ its complex counterpart and the additional momentum
$Q_{_{+}}$ which does not have a direct counterpart in complex quantum mechanics. From these momenta, we get
\begin{equation}
v_{_\pm}\left[\,V_{\0}\,,\,|W_{\0}\,|\,\right]=\Big(\frac{\partial Q_{_{\pm}}}{\partial E}\Big)^{^{-1}}=\frac{Q_{_\pm}}{E}\,\left(\,1\,\pm\,\frac{V_{\0}^{^2}+|W_{\0}|^{^2}}{\sqrt{E^{ ^2}\,V_{\0}^{^2}+(E^{^2}-m^{\2})|W_{\0}|^{^2}}}\,
\right)^{^{-1}}\,\,,
\label{v_pm}
\end{equation}
It is interesting to observe that, in the complex  limit, $|W_{\0}|\to 0$, we find
\begin{equation}
v_{_-}\left[\,-\,V_{\0}\,,\,0\,\right]=v_{_+}\left[\,V_{\0}\,,\,0\,\right]\,\,.
\end{equation}
This means that the $V_{\0}$-antiparticles solution in quaternionic quantum mechanics comes directly from the $Q_{_{+}}$ momentum  solution and, consequently, we have not the necessity to use the  Feynman-Stuckelburg rule. This is clear in Fig.\,2, where the contour plots of $v_{_+}^{^{2}}$, for a fixed value of the incoming energy $E$ and for a fixed values of the pure quaternionic potential $|W_{\0}|$, clearly show that the velocity increases for increasing values of the complex potential $V_{\0}$, typical antiparticle behavior.  This conclusion {\em drastically}  changes if we consider the effect of the pure quaternionic potential on the $Q_{_{+}}$ momentum  solutions.
In this case, we find $|W_{\0}|$-particle like solutions. Indeed, the plots of Fig.\,2 show that,  for a fixed value of the incoming energy $E$ and for a fixed values of the complex potential $V_{\0}$,  the velocity decreases for increasing values of the pure quaternionic potential $|W_{\0}|$. Thus, the pure quaternionic potential compensates the complex potential effect  decelerating the $V_{\0}$-antiparticles. In the limit $|W_{\0}|\gg V_{\0}$,
\begin{equation}
\label{eqr1}
v_{_+}\left[\,V_{\0}\,,\,|W_{\0}|\gg V_{\0} \,\right] \approx  \frac{p}{E}\,\,,
\end{equation}
and the propagation in the potential region  is the same of the propagation in free space. The complex potential acceleration is, in this limit, completely removed by the pure quaternionic potential effect.

In the $Q_{_{-}}$ momentum solutions, we first observe the presence of a tunneling energy zone which
corresponds to $Q_{_{-}}^{^{2}}<0$ solutions and consequently to {\em evanescent} solutions, then we note
that pure quaternoinic perturbations on the complex potential $V_{\0}$ reduces such a tunneling energy zone, see Fig.\,3. In the limit $|W_{\0}|\gg V_{\0}$,
\begin{equation}
\label{eqr2}
v_{_-}\left[\,V_{\0}\,,\,|W_{\0}|\gg V_{\0} \,\right] \approx -\, \frac{p}{E}\,\,.
\end{equation}
This means that by increasing  the pure quaternionic potential we can  remove the $V_{\0}$-antiparticle acceleration.

The analysis done in this section represents only a first discussion on the
properties  of the quaternionic solutions.  The dynamical conditions in which this kind
of solution appears  have to be investigated  in detail  to see  how quaternionic potentials  modify the standard complex scattering.

\section{\normalsize QUATERNIONIC DIRAC SOLUTIONS}\label{s5}

In this section, we give the explicit solution for the Dirac quaternionic spinors and discuss two limit cases, i.e. the complex and the pure quaternionic limit.\\

\noindent $\bullet$ $\mathbf{Q_{_{-}}}$ {\bf \small SPINOR SOLUTION}\\

\noindent The spinorial wave-function
\[\psi_{_{-}}(z) = (\,u_{_{-}} + j\,w_{_{-}}\,)\,\exp[\,i\,Q_{_{-}}z\,]\,\,,\]
is solution of the quaternionic Dirac equation for quaternionic spinors $u_{_{-}}$ satisfying
\begin{equation}
\left[\,E^{^{2}}+Q_{_{-}}^{^{2}} - m^{\2} - V^{^{2}}_{\0} -  \left|W_{\0}\right|^{^{2}} -2\,m\,V_{\0}\,\beta - 2 \,Q_{_{-}}\,(E+m\,\beta)\,\alpha_{\3}\right] u_{_-}   =  0\,\,,
\end{equation}
see the first of Eqs.\,(\ref{eigen_u}). By choosing the up component as
$\boldsymbol{\chi}=\{\,[\,1\,\,0\,]^{t}\,,\,[\,0\,\,1\,]^{t} \,\}$, from the previous equation, we obtain
\begin{equation}
u_{_{-}} =  \left[\,
\begin{array}{c}
 \boldsymbol{\chi} \\  \\ A_{_{-}}\,\sigma_{\3}\,\boldsymbol{\chi}
 \end{array}
 \right]\,\,,
 \end{equation}
with
 \[A_{_{-}} =\displaystyle{\frac{Q_{_{-}}}{E-V_{\0}+m -\frac{\delta}{E-m}}}\,\,.\]
The spinor  $w_{_{-}}$ can be then found by using the second of Eqs.\,(\ref{emms}),
 \begin{eqnarray}
w_{_{-}} & =  & - \,W_{\0}\, \left(\,E-Q_{_{-}}\,\alpha_{\3}+m\,\beta\,+V_{\0}\right)^{^{-1}}\,u_{_{-}}\nonumber \\
& =  & - \,\frac{W_{\0}}{q_{_{+}}^{^{2}} - Q_{_{-}}^{^{2}} }\, \left(\,E+Q_{_{-}}\,\alpha_{\3}-m\,\beta\,+V_{\0}\right)\,u_{_{-}}\nonumber \\
& = &  - \,W_{\0}\,
 \left[\,
\begin{array}{c}
 M_{_{-}}\,\boldsymbol{\chi} \\  \\ N_{_{-}}\,\sigma_{\3}\,\boldsymbol{\chi}
 \end{array}
 \right]\,\,,
 \end{eqnarray}
 where
 \[
M_{_-}= \frac{Q_{_-}A_{_-} + E-m + V_{\0} }{q_{_+}^{\2} - Q_{_-}^{^{2}}}\hspace*{1cm}\mbox{and}\hspace*{1cm}
N_{_-}= \frac{ (E+m + V_{\0})A_{_-}+ Q_{_-} }{q_{_+}^{\2} - Q_{_-}^{^{2}}}\,\,.
\]
Finally, the Dirac spinorial wave-function in the quaternionic potential region for the $Q_{_{-}}$ momentum solution is given by
\begin{equation}
\psi_{_{-}}(z)= \left[\,
\begin{array}{c}
\left(\,1-j\,W_{\0}\,M_{_{-}}\,\right)\,\boldsymbol{\chi} \\  \\ \left(\,A_{_{-}}-j  \,W_{\0}\,N_{_{-}}\,\right)\,\sigma_{\3}\,\boldsymbol{\chi}
 \end{array}
 \right]\,\exp[\,i\,Q_{_{-}}z\,]\,\,.
\end{equation}

\noindent $\bullet$ $\mathbf{Q_{_{+}}}$ {\bf \small SPINOR SOLUTION}\\

\noindent The spinorial wave function
\[\psi_{_{+}}(z) = (\,u_{_{+}} + j\,w_{_{+}}\,)\,\exp[\,i\,Q_{_{+}}z\,]\,\,,\]
is solution of the quaternionic Dirac equation for spinors $w_{_{+}}$ satisfying
\begin{equation}
\left[\,E^{^{2}}+Q_{_{+}}^{^{2}} - m^{\2} - V_{\0}^{^{2}} -  \left|W_{\0}\right|^{^{2}} -2\,m\,V_{\0}\,\beta - 2 \,Q_{_{+}}\,(E-m\,\beta)\,\alpha_{\3}\right] w_{_+}   = 0\,\,,
\end{equation}
see the second of Eqs.\,(\ref{eigen_u}). The choice of $\boldsymbol{\chi}$ as the down component of   $w_{_{+}}$ implies
\begin{equation}
w_{_{+}} =  \left[\,
\begin{array}{c}
  A_{_{+}}\,\sigma_{\3}\,\boldsymbol{\chi} \\ \\ \boldsymbol{\chi}
 \end{array}
 \right]\,\,,
 \end{equation}
with
 \[A_{_{+}} =\displaystyle{\frac{Q_{_{+}}}{E+V_{\0}+m +\frac{\delta}{E-m}}}\,\,.  \]
The spinor  $u_{_{+}}$ is then obtained  from the first of Eqs.\,(\ref{emms}),
\begin{eqnarray}
u_{_{+}} & =  & - \,W_{\0}^{^{*}}\, \left(\,E-Q_{_{+}}\,\alpha_{\3}-m\,\beta\,-V_{\0}\right)^{^{-1}}\,w_{_{+}}\nonumber \\
& =  & - \,\frac{W_{\0}^{^{*}}}{q_{_{-}}^{^{2}} - Q_{_{+}}^{^{2}} }\, \left(\,E+Q_{_{+}}\,\alpha_{\3}+m\,\beta\,-V_{\0}\right)\,w_{_{+}}\nonumber \\
& = &  - \,W_{\0}^{^{*}}\,
 \left[\,
\begin{array}{c}
 N_{_{+}}\,\sigma_{\3}\,\boldsymbol{\chi}\\ \\  M_{_{+}}\,\boldsymbol{\chi}
 \end{array}
 \right]\,\,,
 \end{eqnarray}
where
 \[
M_{_+}= \frac{Q_{_+}A_{_+} + E-m - V_{\0} }{q_{_-}^{\2} - Q_{_+}^{^{2}}}\hspace*{1cm}\mbox{and}\hspace*{1cm}
N_{_+}= \frac{ (E+m - V_{\0})A_{_-}+ Q_{_+} }{q_{_-}^{\2} - Q_{_+}^{^{2}}}\,\,.
\]
The Dirac spinorial wave-function in the quaternionic potential region for the $Q_{_{+}}$ momentum solution then reads
\begin{equation}
\psi_{_{+}}(z)= \left[\,
\begin{array}{c}
 \left(\,- \,W_{\0}^{^{*}}\,N_{_{+}}+j\,A_{_{+}}\,\right)\,\sigma_{\3}\,\boldsymbol{\chi}
 \\ \\ \left(\,-\,W_{\0}^{^{*}}\,M_{_{+}}+j\,\right)\,\boldsymbol{\chi}
 \end{array}
 \right]\,\exp[\,i\,Q_{_{+}}z\,]\,\,.
\end{equation}

\noindent $\bullet$ {\bf \small COMPLEX LIMIT}\\

\noindent For $W_{\0}\to 0$, we find
\begin{equation}
\psi_{_{-}}(z)\,\,\,\to\,\,\, \left[\,
\begin{array}{c}
\boldsymbol{\chi} \\  \\ \displaystyle{\frac{q_{_{-}}}{E-V_{\0}+m}}\,\,\sigma_{\3}\,\boldsymbol{\chi}
 \end{array}
 \right]\,\exp[\,i\,q_{_{-}}z\,]\,\,,
 \end{equation}
 and
 \begin{equation}
\psi_{_{+}}(z)\,\,\,\to\,\,\, j\,\left[\,
\begin{array}{c}
 \displaystyle{\frac{q_{_{+}}}{E+V_{\0}+m}}\,\,\sigma_{\3}\,\boldsymbol{\chi}
 \\ \\ \boldsymbol{\chi}
 \end{array}
 \right]\,\exp[\,i\,q_{_{+}}z\,]\,\,.
\end{equation}

\noindent $\bullet$ {\bf \small PURE QUATERNIONIC LIMIT}\\

\noindent For $V_{\0}\to 0$, by considering $p>|W_{\0}|$, we have
\[ Q_{_{\pm}}\,\,\,\to\,\,\, p \,\,\pm\, |W_{\0}|\,\,,\,\,\,\,\,\,\,\,
A_{_{\pm}}\,\,\,\to\,\,\,\frac{p}{E+m}\,\,,\,\,\,\,\,\,\,\,
M_{_{\pm}}\,\,\,\to\,\,\,\mp\,\frac{1}{|W_{\0}|}\,\frac{p}{E+m}\,\,,\,\,\,\,\,\,\,\,
N_{_{\pm}}\,\,\,\to\,\,\,\mp\,\frac{1}{|W_{\0}|}\,\,.
\]
Consequently,
\begin{equation}
\psi_{_{-}}(z)\,\,\,\to\,\,\, \left[\,
\begin{array}{c}
\left(\,1-j\,\displaystyle{\frac{W_{\0}}{|W_{\0}|}\,\frac{p}{E+m}}\,\right)\,\boldsymbol{\chi} \\  \\ \left(\,\displaystyle{\frac{p}{E+m}}-j  \,\displaystyle{\frac{W_{\0}}{|W_{\0}|}}\,\right)\,\sigma_{\3}\,\boldsymbol{\chi}
 \end{array}
 \right]\,\exp[\,i\,(p\,-|W_{\0}|)\,z\,]
\end{equation}
and
\begin{equation}
\psi_{_{+}}(z)\,\,\,\to\,\,\, \left[\,
\begin{array}{c}
 \left(\, \displaystyle{\frac{\,W_{\0}^{^{*}}}{|W_{\0}|}}+j\,\displaystyle{\frac{p}{E+m}}\,\right)\,\sigma_{\3}\,\boldsymbol{\chi}
 \\ \\ \left( \,\displaystyle{\frac{\,W_{\0}^{^{*}}}{|W_{\0}|}\,\frac{p}{E+m}}+j\,\right)\,\boldsymbol{\chi}
 \end{array}
 \right]\,\exp[\,i\,(p\,+|W_{\0}|)\,z\,]\,\,.
\end{equation}

\section{\normalsize CONCLUSIONS}\label{s6}

In this article, we have studied the spinorial solution of  Dirac particles  and
anti-particles in the  quaternionic potential region. The  two quaternionic solutions $\psi_{_-}$ and $\psi_{_+}$, corresponding to the momenta  $Q_{_-}$ and $Q_{_+}$, describe the behavior of Dirac spinors in the potential region. A first interpretation, given by analyzing the energy zones, is that antiparticle can be described by the $\psi_{_+}$ solutions and that the pure quaternionic potential acts  on such solutions as a decelerating potential, see Fig.\,2. The $\psi_{_-}$ solution represents the quaternionic generalization of the complex spinorial solution. In the diffusion zone, such a solution represents  particles. Indeed, the complex part of the potential acts decelerating $\psi_{_-}$ solutions. In the diffusion zone, the pure quaternionic potential acts on such a solution in an ambiguous way, it can accelerate or decelerate particles, see Fig.\,3.  This novel and intriguing behavior needs to  be further
investigated in dynamical scattering problems.


The results obtained in this paper represents an  initial effort towards
a complete understanding of quaternionic Dirac particles. The solutions, obtained in the quaternionic
potential region, have to be matched at the discontinuity of the potential. The matching conditions  lead to the calculation of the  reflection and transmission coefficients and this allows to determine the dynamical perturbations generated by  quaternionic potentials. In this spirit, it will be also interesting to extend our analysis from plane waves to wave packets.

We conclude this paper observing that pure quaternionic potentials imply drastic changes in the velocity of Dirac particles.
For example from  Eq.\,(\ref{eqr1}), obtained in  the limit $|W_{\0}|\gg V_{\0}$, we find that particles move in pure quaternionic regions with the same velocity of free particles. The propagation throughout a pure quaternionic barrier surely represents an interesting case to see an anomalous  behavior with respect to the standard theory. Indeed, by solving step by step the barrier problem, we expect that the transit time throughout a pure quaternionic barrier is the same of free propagation. Nevertheless, with respect to free propagation the quaternionic solution  present and additional phase and this phase  could be responsible for new interference effects. This brief concluding discussion represents for us a motivation to begin a detailed study of the transmission of wave packet throughout pure quaternionic barriers.

\vspace{.8cm}

\noindent
\textbf{\footnotesize ACKNOWLEDGEMENTS}\\
The authors gratefully thank  the CNPq (Stefano De Leo) and CAPES (Sergio Giardino)  for  the financial support and an anonymous referee for his very useful suggestions and observations. We also thank the referee for having stimulated the final discussion presented in our conclusions.

\pagebreak
\begin{figure}
\hspace*{-1.6cm}
\epsfig{file=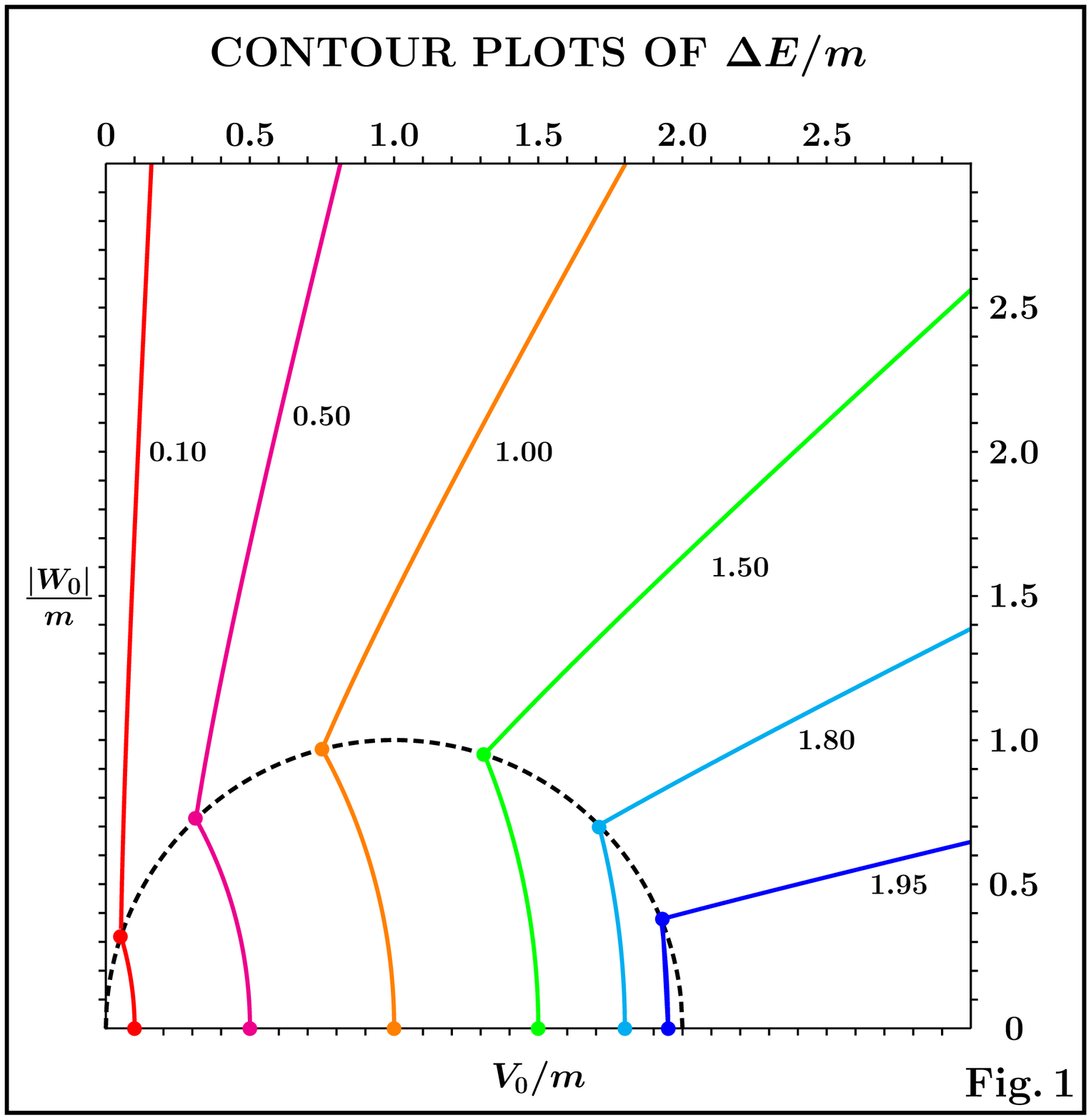,width=19cm}
\vspace*{-7.5cm}
\caption[F1]{ Contour plots of the tunneling energy zone as function of
  the complex, $V_{\0}$, and pure quaternionic,
  $|W_{\0}|$, potentials. The dashed line represents the curve $|W_{\0}|^{^2}+(V_{\0}-m)^{^2}=m^{\2}$.}
\end{figure}
\pagebreak

\begin{figure}
\hspace*{-1.6cm}
\epsfig{file=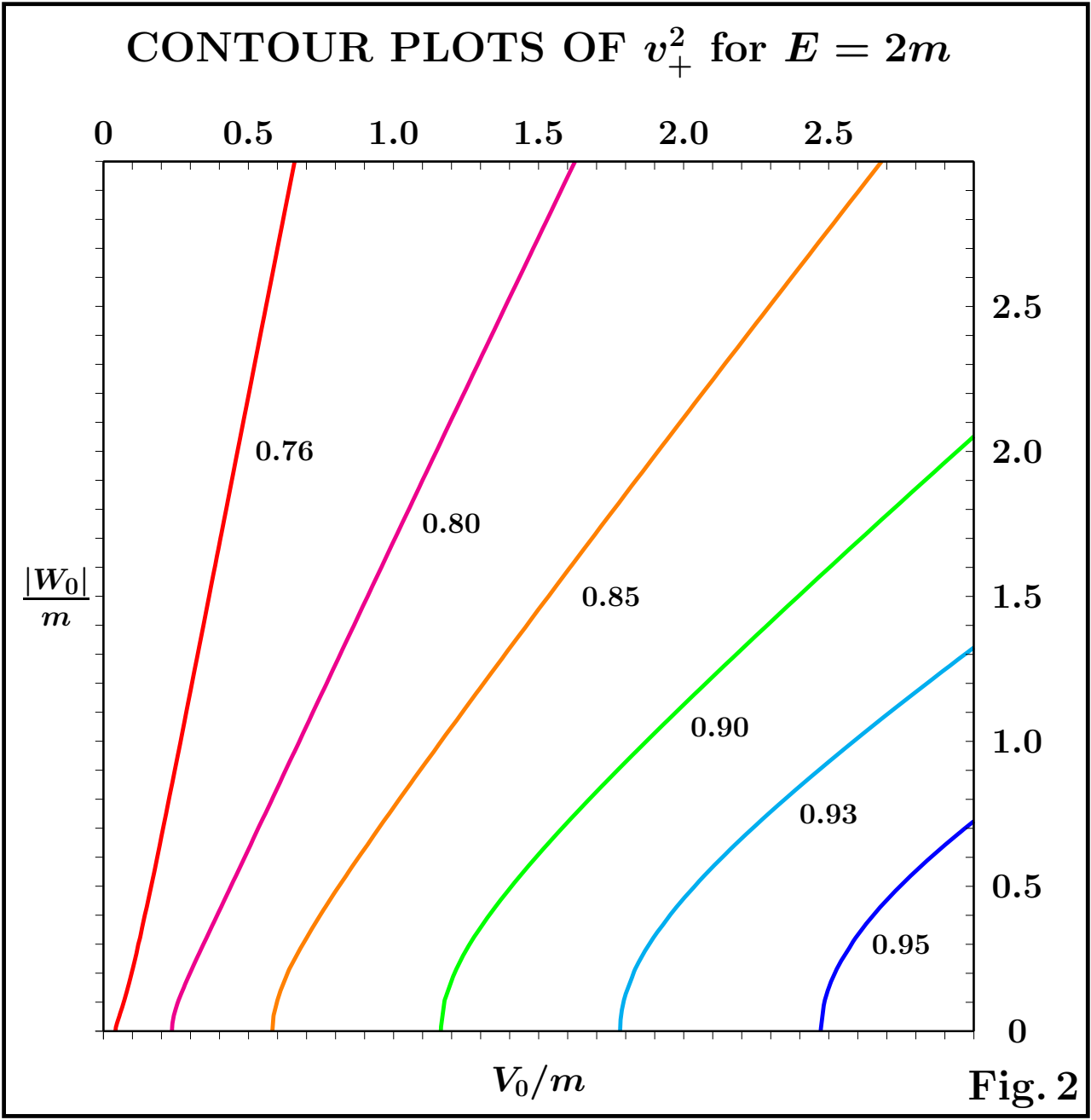,width=19cm}
\vspace*{-7.5cm}
\caption[F1]{The $V_{\0}$-antiparticle like squared velocity, $v_{_+}^{\2}$, is plotted, for a fixed incoming energy, as  function of
  the complex, $V_{\0}$, and pure quaternionic,
  $|W_{\0}|$, potentials. For a fixed $|W_{\0}|$, the complex potential $V_{\0}$ acts as an accelerating potential (anti-particle behavior). The pure quaternionic potential, for a fixed $V_{\0}$, removes this acceleration.}
\end{figure}

\pagebreak
\begin{figure}
\hspace*{-1.6cm}
\epsfig{file=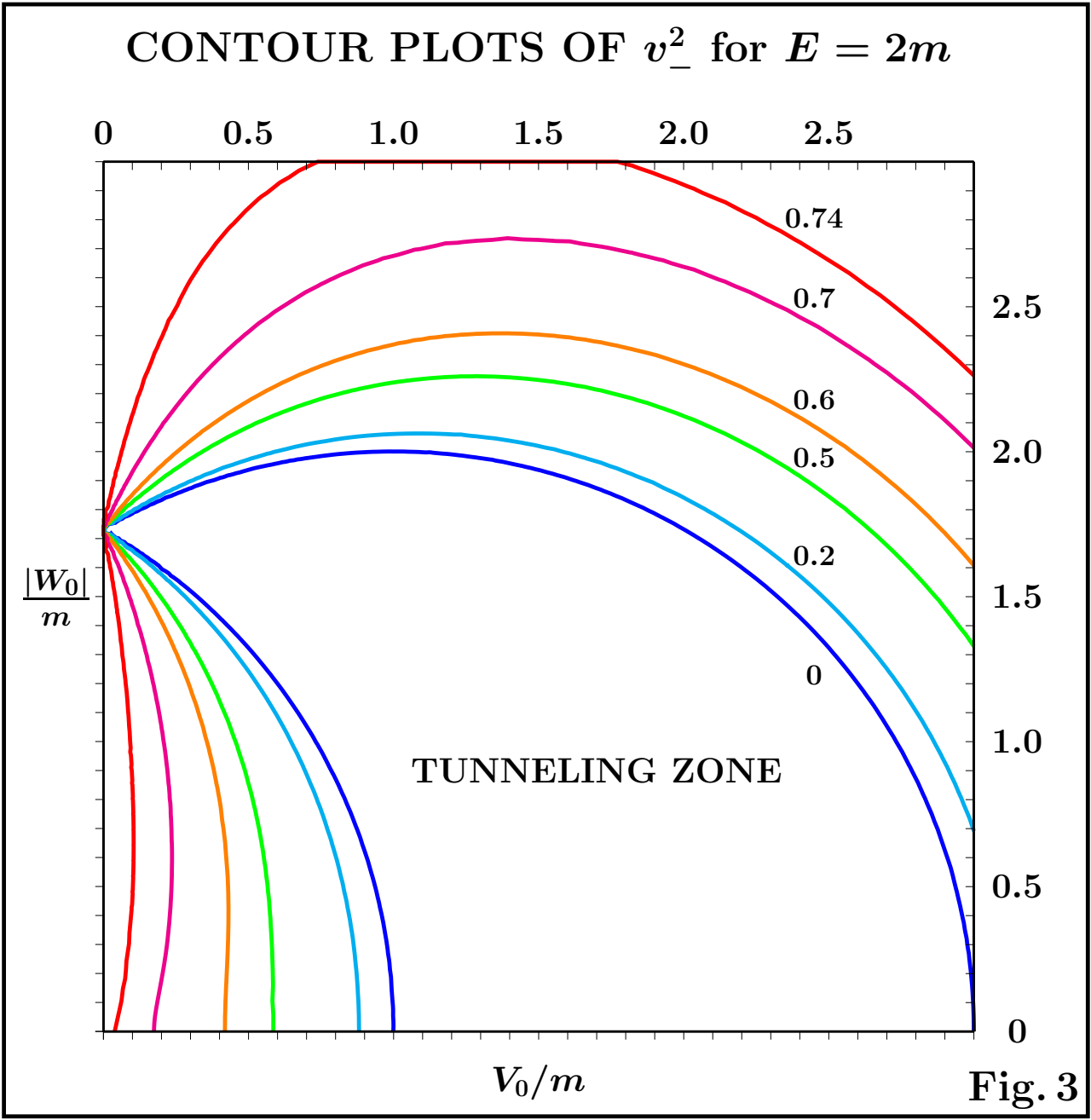,width=19cm}
\vspace*{-7.5cm}
\caption[F2]{The region between the $v_{_-}^{\2}=0$ lines represents the tunneling
  zone. In the zone below the tunneling zone, the complex potential acts as a decelerating potential (particle behavior).  For pure quaternionic potential ($V_{\0}=0$), we recover the velocity in free space tends ($p^{\2}/E^{^{2}}=3/4$, for an incoming energy of $2\,m$).}
\end{figure}

\end{document}